# Emergent Atomic Environments in Twisted Bilayer Graphene and Their Use in the Prediction of the Vibrational Properties

Dilara Ickecan[1,2*], Yunus Emre Okyayli[3], Erdi Ata Bleda[1], Dogan Erbahar [4]

[1]*Marmara University, Dept. of Physics, 34722, Ziverbey, Istanbul, Türkiye*
[2]*Istanbul Ticaret University, Dept. of Computer Programming, 34840, Istanbul, Türkiye*
[3]*Dogus University, Dept. of Mechanical Engineering, Erbahar Researcher Lab, Umraniye, 34775, Istanbul, Türkiye*
[4]*Dogus University, Faculty of Engineering, Dept. of Mechanical Engineering, Umraniye, 34775, Istanbul, Türkiye*

## ABSTRACT

While Bernal stacked bilayer graphene bears two distinct atom types in its lattice, there exists no analytical framework addressing the number of atomic environments that emerge in twisted bilayer graphene superlattices. Here, we computationally analyze 140 different twisted bilayer superlattices using descriptor functions to study the emergent local environments. Our study reveals that the number of atoms with unique local environments depend on the superlattice size linearly manifesting itself on two distinct lines in accordance with the respective space groups. We then introduce the use of local environments in the investigation of vibrational properties. Local phonon density of states of the atoms with unique local environments are calculated by molecular dynamics simulations and are used to train a machine learning model. This model is used to predict the phonon density of states of twisted bilayer structures. Performance of the trained model is cross validated and discussed thoroughly via different selection of training and test sets. It is shown that the model proves effective in predicting the vibrational properties of any given twisted bilayer structure. Since the generic method presented reaches far beyond twisted bilayer graphene the possible applications ranging from non-periodic structures to strain induced moiré superlattices are also discussed.

## 1. Introduction

The moiré patterns observed in twisted bilayer graphene (TBLG) structures are optical fringe-like features that emerge as a result of rotation of layers with respect to each other around their common normal axis. A new periodicity is generally created that manifests itself in a supercell which bears a related rotational symmetry of the underlying lattices. The mathematical properties of these rotational moiré lattices have gained interest following the observations of some novel physical phenomena in twisted structures. The studies on moiré lattices mainly focus on topics such as nomenclature and classification schemes [1, 2], creation conditions [3] as well as relationships between some characteristic parameters like rotation angle [3], lattice mismatch [4, 5] and superlattice size. Most of the theoretical studies regarding the electronic and vibrational properties of twisted systems look at the effect of rotation and folding on the reciprocal lattice to interpret the modifications of physical properties via the relative band structures [6-8].

[*]Corresponding author.
E-mail address: dickecan@ticaret.edu.tr (Dilara Ickecan).

.

All extensive studies on TBLG begin with the introduction of an enumeration scheme of the possible structures. This can be achieved via specifying a commensurability condition that would give a superlattice. Alternative ways of deriving this condition (which are all equivalent) have been reported in literature [1-3]. Here we use the convention introduced by Artaud et al. [1] to enumerate the first few thousand lattices with increasing superlattice parameter. The creation of the structures were performed by first generating a list of index pairs $(m, n)$ for the graphene's hexagonal lattice points. This list was then used to create a list of vector lengths by the formula $\sqrt{m^2 + n^2 - mn}$ which was reported in Ref [1]. The commensurability condition is then defined as one of these vector lengths to be an integer multiple of the other since this means they can be overlapped by a rotation. The angle between these vectors then would be computed and one of the lattices would be rotated with respect to the other to create the moiré supercell.

Fig. 1 demonstrates the relationship of lattice parameters of these lattices with respect to their relative rotation angles. The red curve represents the lower bound for the lattice parameter (L) for a given rotation angle $(\theta)$ and is drawn from the analytical expression $L = 1.2295\ \text{Å}/\sin\left(\frac{\theta}{2}\right)$. This formula can be derived from the previously reported [9] "lower bound" expression between the number of atoms in the supercell (N) and the rotation angle $(\theta)$: $N = 1/\sin^2\left(\frac{\theta}{2}\right)$ by considering the areal atom density of bilayer graphene. A general analytical expression of the relationship between the rotation angle and the lattice parameter "seems" to have a decaying trend however, an exact expression regarding this behavior has been evasive from the current scientific literature.

There are also studies which can incorporate the effect of lattice mismatch on lattice size in the case of two different layers [4, 5]. These studies all focus on the supercell itself characterizing the size, indexing, orientation, etc. and say nothing about the local features appearing in each lattice.

.

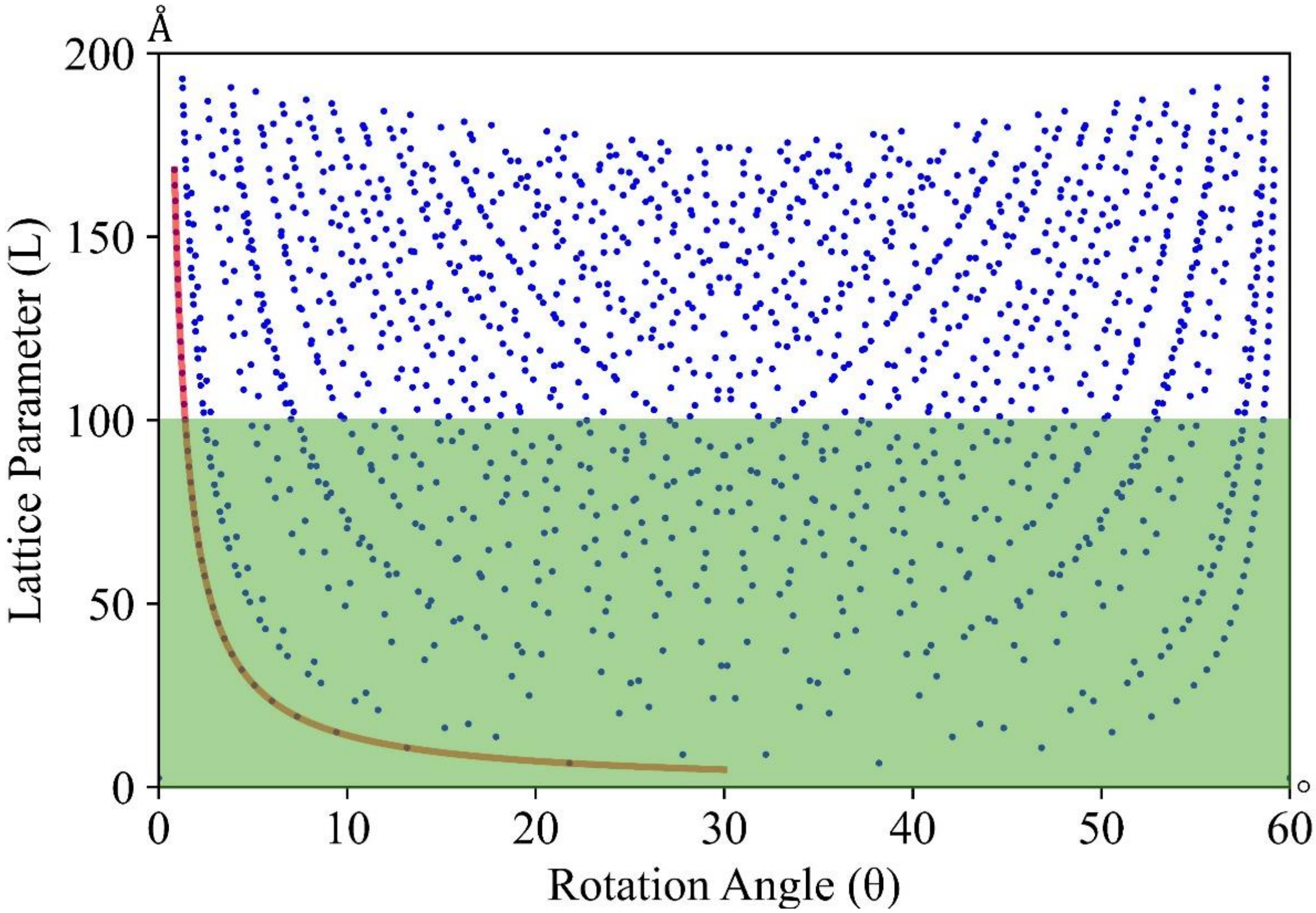

Fig. 1. The dependence of lattice parameter on the twisting angle for 1474 moiré lattices appearing in TBLG. Each blue dot represents a moiré lattice with a unique rotation angle. The mirror symmetry of the plot around $\theta = 30°$ line stems from the inherent rotational symmetries of the structures. Red curve is the lower bound on the lattice parameter. The structures investigated in this study are the ones which have lattice parameter less than 100 Å (green area).

However, we believe that in physical terms, the characterization of moiré superlattices by their local environments is crucial in making sense of their physical properties. In this regard, the emergent local environments of these superlattices can as well constitute a basis of a classification scheme. In this study, we survey this uncharted territory towards an understanding of the sub patterns and local environments as well as their effects on the physical properties. We limit this study to the structures with lattice parameters less than 100 Å's which constitute 140 structures.

## 2. Identification of local environments

First, it is important to note the empirical observation that the number of different atomic environments emerging in the superlattice is unique to the corresponding moiré pattern. To understand what is meant by this, let us initially consider Bernal stacked bilayer graphene. This structure has two different "types" of atoms with respect to their local environments: one sitting exactly on top of an atom on the opposite layer and the second sitting in the midpoint of a hexagon of the opposite layer. The upper panel of Fig. 2(a) illustrates this by using different colors to distinguish the two types (Throughout this manuscript the expression "type" is used to refer to carbon atoms with a unique local atomic environment).

.

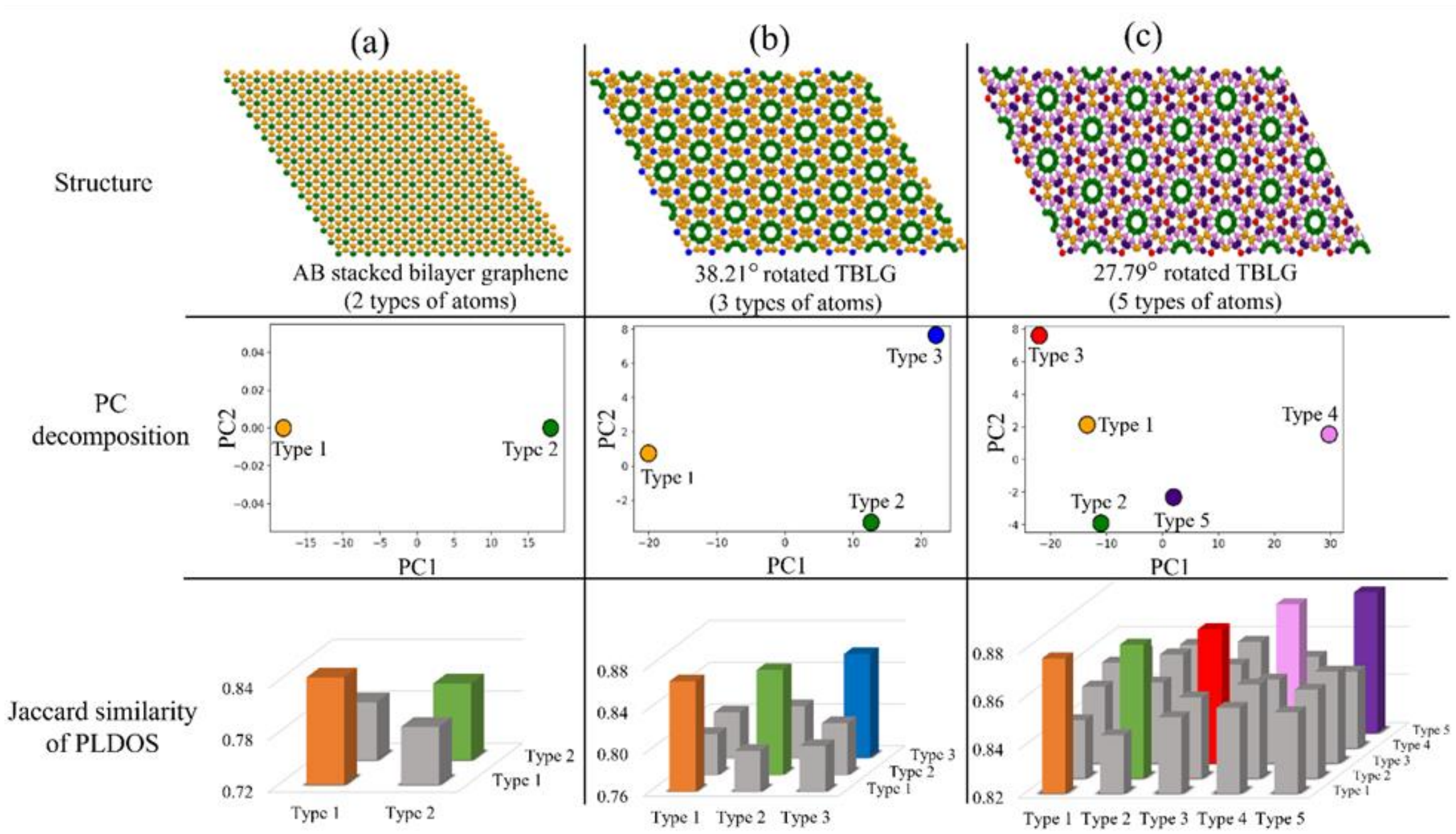

Fig. 2. In the upper panel atoms are coloured according to their local environments in (a) Bernal stacked (b) 38.21° rotated and (c) 27.79° rotated TBLG's. Middle panel demonstrates the clustering of atoms with different local environments achieved by principal component analysis of SOAP descriptors. Lower panel shows the Jaccard similarity indices of phonon local density of states emerging from different types of atoms in the structure compared with respect to each other.

Now, turning our attention to TBLG, one realizes that the situation immediately becomes non-trivial if not chaotic. A relatively manageable case is shown in the upper panel of Fig. 2(b) for the demonstration of the idea. Here, a moiré superlattice with a rotation angle of 38.21° is shown. One would identify 3 different types of atoms only after a careful visual inspection if they were not colored differently as in the Figure. 27.79° rotated TBLG in the upper panel of Fig. 2(c) stands out yet as a more complicated case where it becomes significantly harder to identify the 5 different types of atoms from the perspective of human pattern recognition. The determination of the number of different types of atoms in a general TBLG supercell is the first mathematical curiosity that we focus on, to which to our knowledge, there exists no explicit analytical solution.

At this point, it is important to note that since supercells are constructed by imposing a translational symmetry, each of them should theoretically belong to a specific space group. There exist 17 space groups in 2-dimensions also known as the wallpaper groups. The space groups that can appear in a TBLG superlattice have been investigated via sublattice parity [10, 11] and it was shown that TBLG lattices can only exhibit two different space groups depending on their respective commensurability conditions. These groups are denoted as *p31m* (group 15) and *p6m* (group 17) in IUC plane group notation [12] and belong to different point group classes by having 3-fold and 6-fold rotational symmetries respectively. While a structure's space group determines the symmetry conditions that should be satisfied by its unit cell with respect to its neighbors it says little about the different local environments that can emerge inside a cell.

.

### *2.1 Method*

Evaluating the similarity or dissimilarity of crystalline environments in an automated framework has been a rapidly developing research endeavor in condensed matter physics since the last 10 years. In this context, a solid quantitative similarity measure not only allows classification of structures but also stands as a crucial step in developing algorithms which can navigate through the configuration space and most importantly allow machine-learning techniques effectively find application in condensed matter studies [13]. The fundamental aspect of this process is the representation of the structure. A good representation is expected to remain invariant with respect to certain symmetries that preserve physical quantities. Yet, it should be able to distinguish between structures and environments effectively in a computationally efficient way. While several approaches have been developed for this purpose in recent years, one particular framework called Smooth Overlap of Atomic Positions (SOAP) [14] stood out as an elegant method amongst others [14-16]. Its ability to capture rotation, translation and permutation invariant atomic environments make it suitable to be applicable to the problem of identifying the emergent local environments of TBLG's.

SOAP uses Gaussian functions centered at the atomic positions to represent the atomic densities. The local environment is then described by the overlap of these atomic densities within a defined cutoff radius. The descriptor vector is generated by a power spectrum, which is calculated by the expansion of atomic densities in the basis of real-valued spherical harmonics. The power spectrum is invariant under rotations and permutations of identical atoms, ensuring that the descriptor captures the local environment robustly. The size of this descriptor depends on the number of terms used in the expansion of the atomic densities.

In this work, SOAP descriptors were used as implemented in DScribe package [17]. The SOAP output in this implementation is given by the partial power spectrum vector $\vec{p}$ where the elements are given as:

$$p_{nn'l}^{Z_1 Z_2} = \pi \sqrt{\frac{8}{2l+1}} \sum_m c_{nlm}^{Z_1\ *} c_{n'lm}^{Z_2}$$

Here $n$ and $n'$ denote the indices for the different radial basis functions up to $n_{max}$ while $l$ denotes the angular degree of the spherical harmonics which extend up to $l_{max}$. $Z_1$ and $Z_2$ denote atomic species.

The coefficients $c_{nlm}^{Z}$ are defined as the following inner products:

$$c_{nlm}^{Z} = \iiint_{R^3} dV g_n(r) Y_{lm}(\theta,\phi) \rho^z(r)$$

where $\rho^z(r)$ is the gaussian smoothed atomic density for atoms with atomic number $Z$ defined as

$$\rho^z(r) = \sum_i^{|Z_i|} e^{-|r-R_i|^2 / 2\sigma^2}$$

$Y_{lm}(\theta,\phi)$ are the real spherical harmonics, and $g_n(r)$ is the radial basis function.

.

For the radial degree of freedom, the selection of the basis functions is not trivial, and multiple approaches may be used. We have chosen spherical gaussian type orbitals as radial basis functions [18], as they allow much faster analytic computation. The optimization of parameters such as $n_{max}$, $l_{max}$, and $\sigma$ are typically performed to fit the needs regarding a target system however, a well optimized set was previously shown to be generic enough to be able to classify various states of carbon [19]. In this study, they are chosen as 10, 10 and 0.5 Å respectively. The optimization is done with respect to both classification and learning performance with little sensitivity to the selection of these parameters. We observe however, that the classification performance is mainly affected by the choice of cutoff radius which is chosen as 30 Å as this value embraced at least half of the lattice parameter of the largest supercell used. In a TBLG lattice, the atomic environment for a given atom is obviously identical with respect to its counterparts in the same layer. The difference comes from the relative locations and orientations of the atoms in the opposite layer. To exploit this fact to full extent, in the classification step we reduce the interlayer distance to 1 Å to enhance the contribution of the opposite layer and this procedure significantly improves the performance of the identification of unique local environments[†].

Middle panels of Fig. 2 (a), (b) and (c) demonstrate the ability of SOAP descriptors to verify the number of types in the corresponding structures via Principal Component Analyses (PCAs). PCA is a linear dimensionality reduction procedure frequently used in classification applications that are prone to the curse of dimensionality. The data is represented on a new orthogonal basis constructed to capture the maximum variance. This is typically monitored using a scree plot to decide how many principal components to retain in the representation of the data. The scree plot calculated by considering all the structures is given in Fig. S1 of Supplementary Materials. These principal components are calculated from descriptor matrices by exploring an orthogonal basis set where the greatest variances in the data are explained.

In this study, first 6 principal components were included in machine learning algorithms which were able to capture the 95% of the observed variance with the above supplied $\sigma$ and cutoff parameters. Convergence tests are performed and a resolution of three decimal places is found to be a sufficient tolerance for separating different types in bigger structures.

### *2.2 Unique local environments as a function of lattice size and space group*

By applying this method to 140 different moiré lattices, 34394 different types of atoms with different atomic environments are identified. The number of emergent types vs. the total number of atoms in each moiré supercell is presented in Fig. 3. The most significant feature of this plot is the grouping of 140 structures on two distinct lines. The $R^2$ values of the linear best fits are both greater than 0.99, which strongly supports the proposed linear model. The effect of space group on the number of emergent local environment is obvious as *p31m* and *p6m* structures lie on different lines. The reciprocal of slopes of *p6m* and *p31m* lines are found to be 36.01 and 5.99 clearly pointing to

[†] At this point, to avoid any confusion we would like to emphasize that the reduction of the interlayer spacing was only used to enhance the performance of SOAP descriptors. In MD simulations they were initially set as 3.44 Å which is the interlayer separation for bilayer graphene.

.

$6^2$ and 6. This finding can be interpreted by referring to the respective lower and higher symmetry of *p31m* and *p6m* groups which leads to significantly higher number of different local environment breeding observed in *p31m* group structures.

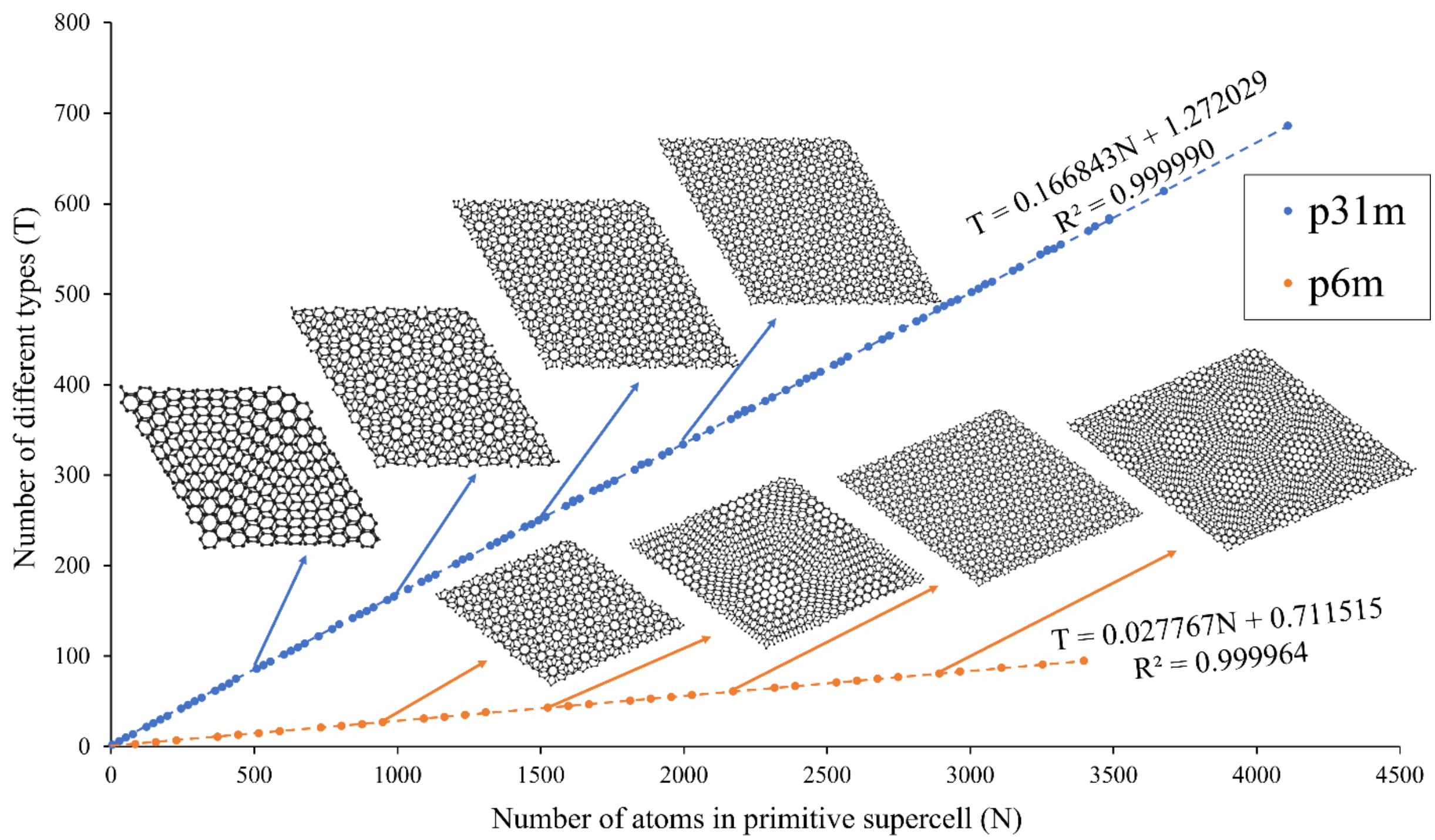

Fig. 3. The dependence of emerging types on the size of the supercell. The 140 investigated structures lie on two different nearly perfect lines. The slopes of the best fit lines are calculated as 0.167 and 0.0278 which correspond to 5.99 and 36.01 atoms in the cell per type respectively. Some randomly selected structures are also shown to point out the lack of any obvious visual pattern regarding the difference between two sets.

The most important point we would like to emphasize, however, is the fact that none of the structures lying on the lines bears exactly 6 nor 36 $(N/T)$ ratio. This is due to unique sub patterns emerging in each structure where local symmetries and occasional overlaps contribute to the total number with different weights. The structures shown in the upper panel of Fig. 2(a), (b) and (c) have for example the respective ratios being 2, 28 and 31.2. A complete list of the structures with their lattice parameters, twist angles, $(N/T)$ ratios as well as space groups are given in the Table S1 of supplementary materials.

These two lines might be regarded as an indication of a new classification scheme of TBLGs with respect to their atomic environments. Some selected structures on both lines are also shown in Fig. 3 to point out that there are no obvious visual clues or human recognizable patterns to indicate the class to which a given structure belongs.

## 3. Use of local environments in predicting vibrational spectra

The significance of the local environment analysis presented above becomes more pronounced when it finds applications in certain domains. In this regard, the problem we focus on is the

.

prediction of the vibrational characteristics of any given TBLG. We approach this problem by decomposing the spectrum into contributions from atoms with unique local environments.

### *3.1 Phonons in TBLG.*

The intense focus on TBLG phonons has been triggered by the discovery of superconductivity in a “magic-angle” rotated sample in 2018 [20, 21]. While this discovery initiated the birth of *twistronics*, the nature of the superconductivity has been under debate since then with no decisive consensus [22-25]. Some studies attribute the superconductivity purely to unique electronic properties [26-29] while others refer to more conventional phonon mediated behavior [30-33]. Hence, the vibrational characteristics of twisted bilayer structures have been one of the major research foci in the field [33-36].

Superconductivity, however, is not the only reason for the interest in TBLG phonons since a complete understanding of the vibrational characteristics would have important implications for low-energy transport and collective phenomena [33]. Indeed, the analysis of the phonon spectra of TBLG has revealed soliton signatures, indicating the complex nature of phonon behavior in this material [37]. Moreover, layer-breathing phonons are recently shown to strongly couple to resonantly excited electrons in TBLG, with their resonance conditions being tunable by the twist angle [38].

First theoretical studies on TBLG phonons predicted that only minor changes would be expected to take place in low frequencies; however experimental results demonstrated a much richer picture throughout the spectrum. For example, G-band which is known to correlate with electron-phonon coupling was shown [39, 40] to change its half-width in near magic angles. While this was related to the emergence of flat bands, some experimental studies have also reported the appearance of new peaks in the Raman spectrum of TBLG for various twisting angles [41-44]. The theoretical and computational approaches to the topic are diverse. Here, we yet introduce a generic scheme of phonon total density of states (PTDOS) prediction from the phonon local density of states (PLDOS). This prediction is realized by a machine learning model which is trained by PLDOS data coming from the unique local atomic environments.

### *3.2 Calculating phonon spectra by molecular dynamics simulations.*

To produce this data, molecular dynamics (MD) simulations are performed on 140 TBLGs. The simulations are run at 300 K using a Langevin thermostat in an NVE microcanonical ensemble. LAMMPS [46] code is used to run the simulations. The simulation boxes are prepared with a suitable size to include a multiple number of supercells in the in-plane directions using periodic boundary conditions. The out of plane direction size of the box is chosen as 60 Å to completely ignore the interactions between the repeated structures in that direction. Interlayer spacing distance of TBLG’s are initially set as 3.44 Å and the structures were relaxed before starting a simulation. MD timesteps are chosen 1 fs and after an initial 2,000,000 step equilibration run the fluctuation-dissipation theory is used to calculate the dynamical matrices of the studied systems during a total of 20,000,000 steps. This is done by the observation of displacement of atoms [45] as implemented in LAMMPS [46] code and related post processing tools are used to deduce the vibrational properties such as phonon

.

dispersion and phonon density of states. The method is shown to be effective in various 2D and 3D systems [47, 48] previously.

Tersoff potential is used for modelling the interactions between carbon atoms in the same layer, while Lennard-Jones potential is employed for handling the van der Waals interactions between atoms in different layers. We use the parametrization of Ref. [49] for Lenard-Jones potential which is optimized for TBLG systems. M. Wen, et al. [50] gives an extensive review of the potentials used in multilayer graphene simulations. It should be noted that none of the potentials in the literature are perfect for predicting the exact frequencies in phonon spectra. The potential scheme we have chosen was previously used for calculating phonon dispersion [51] of 13.2° and 21.79° degrees rotated TBLG by using a lattice dynamics approach. The mentioned study has demonstrated that TBLG's have a much richer phonon dispersion picture throughout the whole frequency range compared to AB stacked bilayer graphene. Here we study the same 21.79° rotated TBLG using molecular dynamic simulations to get a similar result which is presented in Fig. 4(a) and 4(b). The out of plane modes which are also known as flexural phonons in the literature comprise mainly the lower frequencies of the vibrational spectra [32]. However, the new curves appearing in the frequency regime of in-plane modes emphasizes the importance of the effect of interlayer coupling on intra layer behaviour of phonons. Moreover, branching of these newly emerged dispersion curves well correlates with the number of different atom types in the cell which we calculate as 6 for 21.8° rotated TBLG. It is worth mentioning that this insight is gained from our MD simulations performed at room temperature and was loosely evident in the studies which stayed in the limits of harmonic approximation.

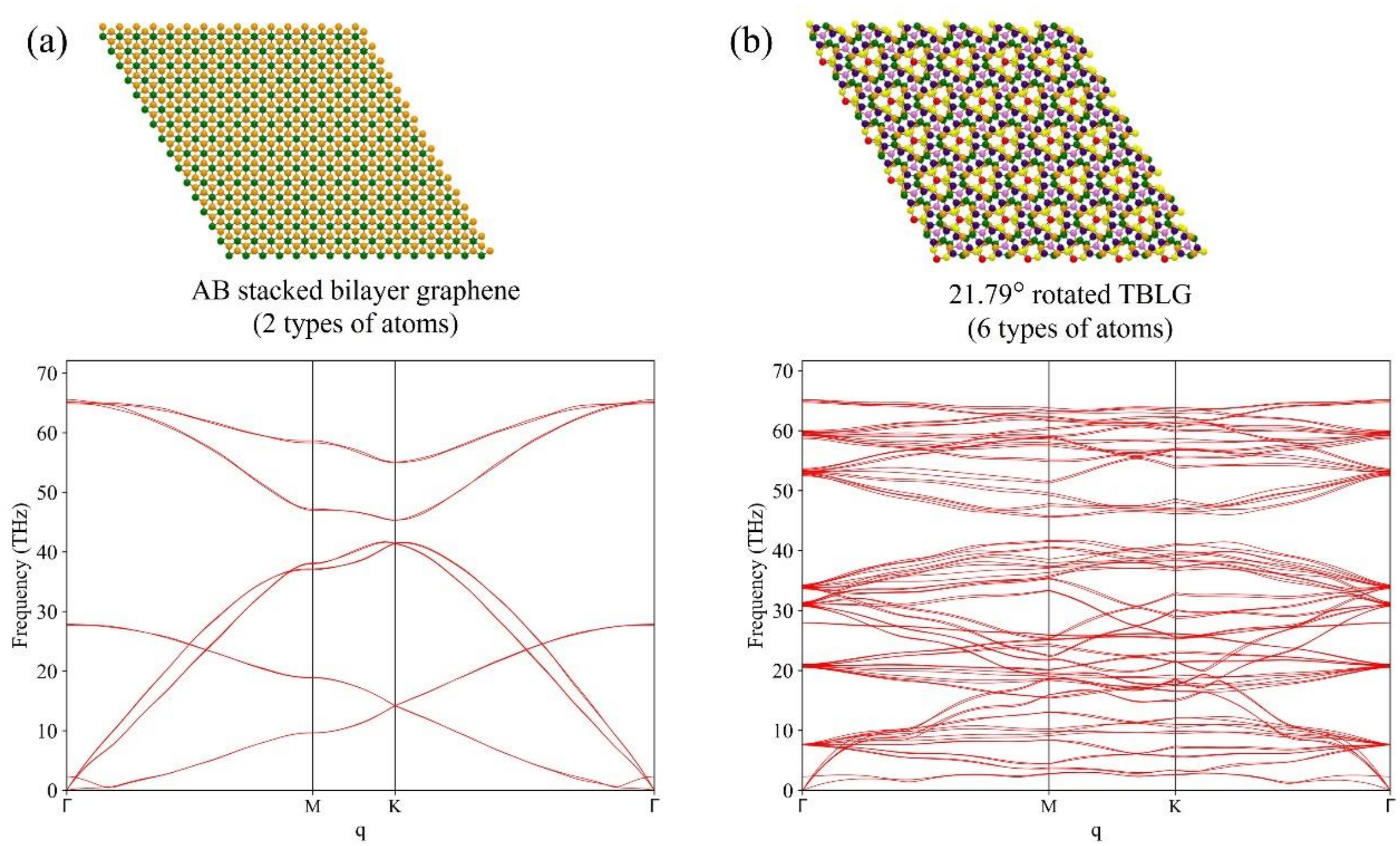

Fig. 4. Phonon dispersion curves calculated from molecular dynamics simulations for (a)AB stacked bilayer graphene and (b) 21.79° TBLG which bears 6 different types of atoms. Besides the breeding of new frequency branches, the effect of local environment on these branches is also evident.

.

The mentioned method with the described potential scheme also gives the relative intensities and shape of the DOS curves sufficiently accurate up to an arbitrary scaling factor in the frequency range. Since it is based on the dynamical observation of motion of atoms, it can also be used to calculate the PLDOS for each atom in the structure, as well as the PTDOS. Obviously, the same types of atoms are expected to contribute to PTDOS equally. However, due to the stochastic nature of our method, an inherent noise is always present in the phonon dispersion curves. To monitor the effect of this on the calculated spectra we perform similarity analysis between the PLDOSs of different types of atoms. Lower panel of Fig. 2(c) shows the results of this analysis for corresponding structures above. Jaccard similarity index [52] is used for this comparison where 10 random atoms from each types are chosen and their PLDOSs are compared with respect to each other. The averages of these similarities are shown in Fig. 2(c). Higher values in the diagonal elements of the matrices verify that different atomic environments would have unique contributions which can be meaningfully characterized above a certain noise level.

### *3.3 Training the model from local environment*

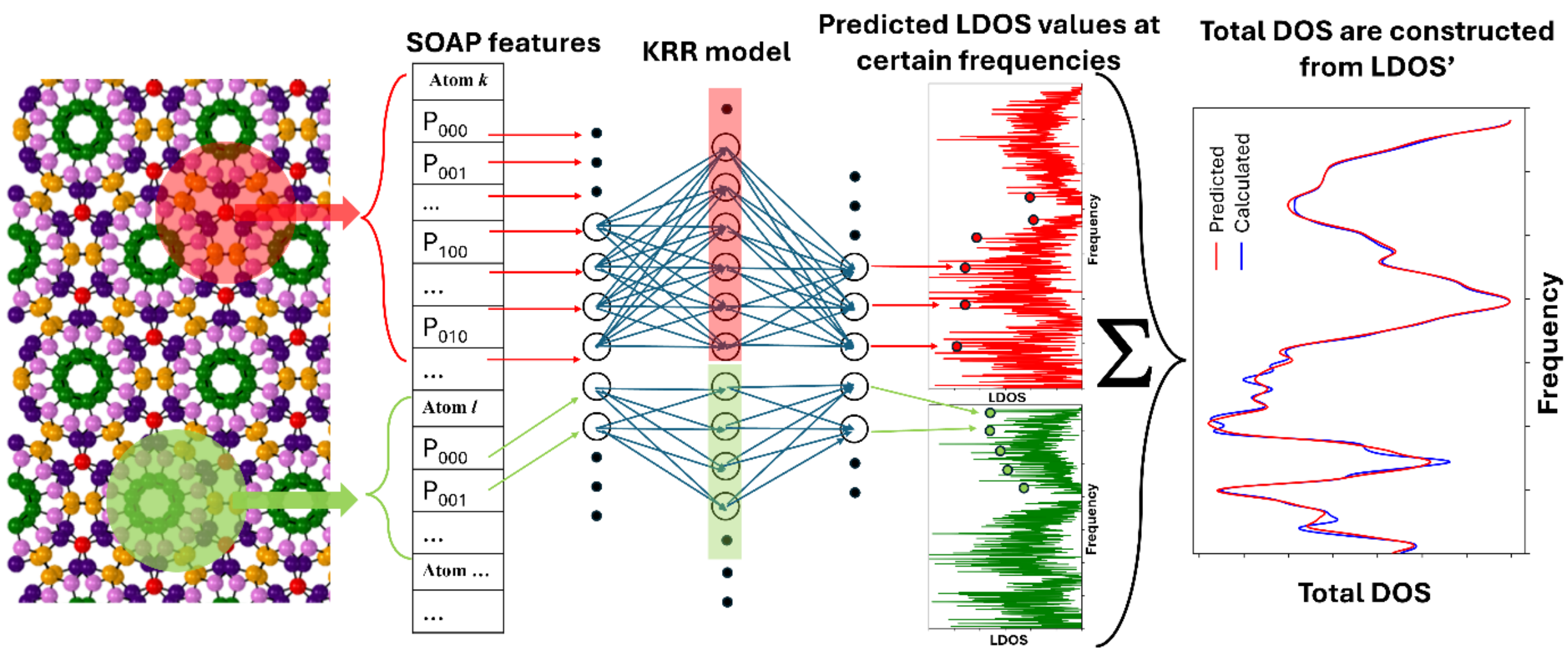

Fig. 5. General scheme of KRR training. 605 SOAP features for each atom with different atomic environments are fed into the input layer. The frequency spectrums are divided into 972 equal parts and PLDOS's are predicted at each frequency. The kernel matrix is trained over approximately 13,500 atomic environments and their respective PLDOS's produced from 112 different TBLG structures. The predicted PLDOS's are used to construct PTDOS of TBLG structures.

Encouraged by this result, we train a Kernel Ridge Regression (KRR) model on taking on the task of predicting the PTDOS spectra of a given TBLG. The general procedure is presented in Fig. 5. The KRR model combines the principles of ridge regression with kernel methods. A kernel function allows for nonlinear regression by mapping input features into a higher dimensional space [53]. The choice of kernel function is critical in KRR model as it determines the shapes of the decision boundary and the model's ability to generalize to unseen data [53]. Further details about the model have been presented in the supplementary material. We examine the performance of three distinct

.

kernel functions: Linear, Laplacian and Radial Basis Function (RBF) in this context. Our results demonstrate that the Laplacian kernel is particularly effective for modelling complex data distributions, outperforming both the Linear and RBF kernels in our tests. Although the Laplacian kernel and the linear kernel perform equally well on the test set, the Laplacian kernel performs better on the training set. While RBF kernel showed improved performance over the linear kernel, it still fell short of Laplacian kernel, likely due to its sensitivity to sigma parameter and potential for overfitting. Moreover, the Laplacian kernel is particularly effective for capturing local structures in data due to its exponential decay, making it suitable for tasks where the relationships between data points are expected to be smooth and continuous [54]. Therefore, we present our results using the KRR model with the Laplacian kernel. Scikit-learn [55] library is used to train the model.

Out of 140 TBLG structures, 112 (80%) of them are randomly selected as the training set. More than 27,000 types of atoms are identified in these 112 structures by our characterization scheme introduced in section 2. The model is trained from PLDOSs of 50% of these 27,000 types of atoms. The 28 (20%) structures which are reserved for training set yields approximately 6000 types of atoms. PLDOSs of the test set are predicted and PTDOSs of structures in the test set are constructed by superposing these predicted PLDOSs. The procedure above is repeated five times, each with a different randomization seed, to establish a higher level of confidence in the model's performance and consistency. This procedure is known as random sub-sampling validation which is commonly used in similar studies. The results presented below are obtained from the best performing seed. The results from the other seeds as well as further relevant details about the KRR model have been presented in the supplementary materials of this manuscript.

A typical example for 2.13º rotated TBLG is shown in Fig. 6(a) where a direct calculation from MD simulation is compared with the prediction of the model. The molecular dynamics simulation had to be run on 8 cores of Intel Xeon Gold 5218R @ 2.10 GHz for approximately 2 weeks to be able to capture the PTDOS of the structure while the trained model takes couple of seconds on a single core of the mentioned model to predict the PTDOS. The performance of the model is tested on magic-angle TBLG which is a much bigger system than the ones included in our training set with 11164 atoms. The results of this prediction with the relevant technical details are presented in supplementary materials.

.

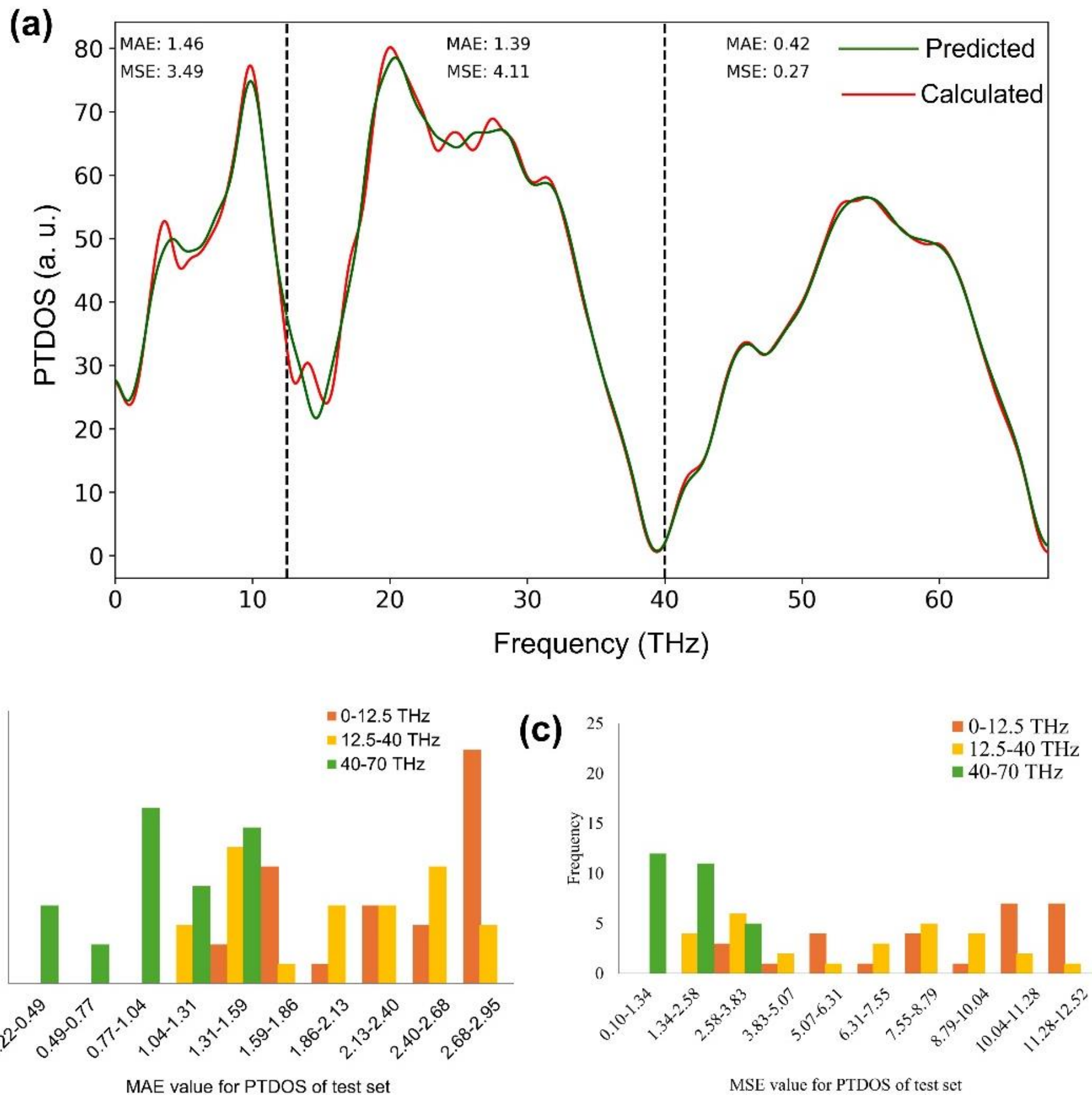

Fig. 6. (a) PTDOS of 2.13° rotated TBLG calculated from MD simulations vs. predicted from trained model. The mean absolute error (MAE) and mean squared error (MSE) values which quantify the accuracy of the prediction are calculated in three regions of the frequency domain. (b) MAE and (c) MSE values for the same frequency domains are presented for 28 TBLG structures used in test set as histograms.

TBLG supercells investigated in the study have lattice parameters ranging from 7 Å to 80 Å. Taking the speed of sound in graphene as a reference, the frequency range of the phonons corresponding to that range has an order of magnitude of 0 to 1 in THz units. Phonons in this regime are expected to be more prone to be coupled to moiré periodicity and hence affected more by the twisting. Indeed, theoretical and experimental studies in literature focus generally on the low frequency end of the spectrum reporting modifications of the phonon spectra to some extent [32-34, 40]. To elaborate on this point, we divide the PTDOS into three regions and separately study the prediction accuracy of our model in each region. Mean squared error (MSE) as well as the mean absolute error (MAE) are used to quantify the prediction accuracy. Fig. 6(a) shows that the mean absolute error (MAE) value points to a higher accuracy towards the higher frequencies where the metric slightly drops in the lower frequency regime. We attribute this to be the result of the richer variety in the lower frequency range of PLDOS where it becomes more challenging for the model to learn and predict. Also, it is important to note that the phonons in this frequency regime have longer wavelengths and hence, the collective behavior of the atoms are more prominent in determining their characteristics. In this aspect, a decomposition scheme which is based on the local features is expected

.

to perform inferior by definition. This is indeed what we observe when we look at all MSE and MAE values for the predicted PTDOS of 28 TBLG structures in the test set which are shown in Fig. 6(b) and Fig. 6(c) as histograms. However, one can see that the accuracy values reported in the long wavelength region are still quite acceptable compared to the established standards in similar machine learning studies. The mean values of the MAE and MSE values of the 28 structures in the test set are reported in Table 1.

**Table 1**. The mean values MAE and MSE values of the structures used in test set.

| Range (THz) | MAE | MSE |
|---|---|---|
| 0 – 12.5 | 2.37 | 8.23 |
| 12.5 – 40 | 1.96 | 6.37 |
| 40 – 70 | 0.99 | 1.61 |

The results of the other four seeds are provided in the supplementary materials of this manuscript to demonstrate the relatively low fluctuation in these statistical indicators.

## 4. Conclusion

In this manuscript, we have addressed the question of identification of emergent atomic environments in TBLG. We have shown that the moiré lattices can be classified with respect to their atomic environments which also correlate with their respective space groups. The problem was beyond the reach of human pattern recognition and lacked an analytical solution. In this regard, we believe that it demonstrates an excellent example of how artificial pattern recognition could complement human insight.

We have also demonstrated that this decomposition approach can be used to analyze and even predict the vibrational properties of TBLG's. Molecular dynamics simulations were employed to calculate vibrational properties. This allowed us not only study much bigger cells than *ab initio* methods could handle but also revealed certain intricate details in the spectrum by capturing the anharmonic effects.

Our main hypothesis was to be able to construct the PTDOS of the structures from PLDOS' of their atoms. This was supported by a similarity analysis and encouraged us to train a machine learning model. Approximately 27000 atoms with different local environments were included in the training set. We have shown that our model could predict the PTDOS of the structures in the test set within the accepted standards in machine learning studies.

The generic method presented here for calculating the vibrational spectra can be generalized to any twisted bi- or multilayered structure as well as structures that bear strain induced moiré patterns. Besides, due to the strong dependence of electronic structure on the local environments, the presented method can promisingly be extended into studies where electronic dispersion, DOS, etc. might be predicted complementarily to the earlier ML studies in this domain. Furthermore, since focusing on the local environment removes the heavy restriction of periodicity which most of the computational methods is constrained to submit, this approach could easily be employed to study non-periodic structures like quasi crystals, amorphous materials, etc.

.

**Declaration of Competing Interest**

The authors declare that they have no known competing financial interests or personal relationships that could have appeared to influence the work reported in this paper.

**Acknowledgements**

The authors would like to thank Dr. Alberto Zobelli and Eftal Gezer for useful discussions at the beginning of the project. We gracefully thank Prof. Savas Berber for the discussions. DE would like to acknowledge financial support from TUBITAK 1001 program (project number: 122F022) and Dogus University scientific research project program (project no: 2020-21-D2-B01). DI would like to thank TUBITAK 2211-C BIDEP program for the scholarship and TUBITAK 1002 program (project number 123F441). EAB would like to thank BAPKO of Marmara University for additional computational support. The simulations were partly performed on TUBITAK-ULAKBIM High Performance and Grid Computing Centre (TRUBA).

## Appendix A. Supplementary data

## Appendix B. A video summary of the work

## Data availability

The scripts used for identifying the local environments, the structure files and sample input files of MD simulations as well as training and testing scripts and a portion of training set can be accessed via the link: https://github.com/Dlrickcn/TBLG_Emergent_Environments

## References

[1] A. Artaud, L. Magaud, T. Le Quang, V. Guisset, P. David, C. Chapelier, J. Coraux, Universal classification of twisted, strained and sheared graphene moiré superlattices, Scientific Reports 6(1) (2016) 25670, https://doi.org/10.1038/srep25670.

[2] M. Ciesler, D. West, S. Zhang, Enumeration of Moire Patterns of a Hexagonal Twisted Bilayer and Intercalated Transition Metals in Twisted h-BN, arXiv preprint arXiv:2211.14476 (2022), https://doi.org/10.48550/arXiv.2211.14476.

[3]M. J. Campanera, G. Savini, I. Suarez-Martinez, M. Heggie, Density functional calculations on the intricacies of Moiré patterns on graphite, Physical Review B 75 (23) (2007) 235449, https://doi.org/10.1103/PhysRevB.75.235449.

[4] S.S. Dindorkar, A.S. Kurade, A.H. Shaikh, Magical moiré patterns in twisted bilayer graphene: A review on recent advances in graphene twistronics, Chemical Physics Impact (2023) 100325, https://doi.org/10.1016/j.chphi.2023.10032.

[5] M. Yankowitz, J. Xue, D. Cormode, J.D. Sanchez-Yamagishi, K. Watanabe, T. Taniguchi, P. Jarillo-Herrero, P. Jacquod, B.J. LeRoy, Emergence of superlattice Dirac points in graphene on hexagonal boron nitride, Nature physics 8(5) (2012) 382-386, https://doi.org/10.1038/nphys2272.

[6] J.-A. Yan, W. Ruan, M. Chou, Phonon dispersions and vibrational properties of monolayer, bilayer, and trilayer graphene: Density-functional perturbation theory, Physical Review B 77 (12) (2008) 125401, https://doi.org/10.1103/PhysRevB.77.125401.

.

[7] N.R. Chebrolu, B.L. Chittari, J. Jung, Flat bands in twisted double bilayer graphene, Physical Review B 99(23) (2019) 235417, https://doi.org/10.1103/PhysRevB.99.235417.

[8] R. Bistritzer, A.H. MacDonald, Moiré bands in twisted double-layer graphene, Proceedings of the National Academy of Sciences 108(30) (2011) 12233-12237, https://doi.org/10.1073/pnas.110817410.

[9] S. Shallcross, S. Sharma, E. Kandelaki, O. Pankratov, Electronic structure of turbostratic graphene, Physical Review B 81(16) (2010) 165105, https://doi.org/10.1103/PhysRevB.81.165105.

[10] L. Zou, H.C. Po, A. Vishwanath, T. Senthil, Band structure of twisted bilayer graphene: Emergent symmetries, commensurate approximants, and Wannier obstructions, Phy.Rev. B 98(8) (2018) 085435, https://doi.org/10.1103/PhysRevB.98.085435.

[11] E.J. Mele, Commensuration and interlayer coherence in twisted bilayer graphene, Physical Review B 81(16) (2010) 161405, https://doi.org/10.1103/PhysRevB.81.161405.

[12] T. Hahn, U. Shmueli, J.W. Arthur, International tables for crystallography (1983).

[13] S. De, A.P. Bartók, G. Csányi, M. Ceriotti, Comparing molecules and solids across structural and alchemical space, Physical Chemistry Chemical Physics 18(20) (2016) 13754-13769,https://doi.org/10.1039/C6CP00415F.

[14] A.P. Bartók, R. Kondor, G. Csányi, On representing chemical environments, Physical Review B 87(18), (2013) 184115, https://doi.org/10.1103/PhysRevB.87.184115.

[15] J. Behler, M. Parrinello, Generalized neural-network representation of high-dimensional potential-energy surfaces, Physical review letters 98(14), (2007) 146401, https://doi.org/10.1103/PhysRevLett.98.146401.

[16] M. Rupp, A. Tkatchenko, K.-R. Müller, O.A. Von Lilienfeld, Fast and accurate modeling of molecular atomization energies with machine learning, Physical review letters 108(5) (2012), 058301 https://doi.org/10.1103/PhysRevLett.108.058301.

[17] J. Laakso, L. Himanen, H. Homm, E.V. Morooka, M.O. Jäger, M. Todorović, P. Rinke, Updates to the DScribe library: New descriptors and derivatives, The Journal of Chemical Physics 158(23) (2023), https://doi.org/10.1063/5.0151031.

[18] M.O. Jäger, E.V. Morooka, F. Federici Canova, L. Himanen, A.S. Foster, Machine learning hydrogen adsorption on nanoclusters through structural descriptors, npj Computational Materials 4(1) (2018), https://doi.org/10.1038/s41524-018-0096-5.

[19] C.B. Mahmoud, A. Anelli, G. Csányi, M. Ceriotti, Learning the electronic density of states in condensed matter, Physical Review B 102(23) (2020) 235130, https://doi.org/10.1103/PhysRevB.102.235130.

[20] Y. Cao, V. Fatemi, A. Demir, S. Fang, S.L. Tomarken, J.Y. Luo, J.D. Sanchez-Yamagishi, K. Watanabe, T. Taniguchi, E. Kaxiras, Correlated insulator behaviour at half-filling in magic-angle graphene superlattices, Nature 556 (2018) 80-84, https://doi.org/10.1038/nature26154.

[21] Y. Cao, V. Fatemi, S. Fang, K. Watanabe, T. Taniguchi, E. Kaxiras, P. Jarillo-Herrero, Unconventional superconductivity in magic-angle graphene superlattices, Nature 556(7699) (2018) 43-50, https://doi.org/10.1038/nature26160.

[22] E. Codecido, Q. Wang, R. Koester, S. Che, H. Tian, R. Lv, S. Tran, K. Watanabe, T. Taniguchi, F. Zhang, Correlated insulating and superconducting states in twisted bilayer graphene below the magic angle, Science Advances 5(9) (2019), https://doi.org/10.1126/sciadv.aaw9770.

[23] Y. Cao, D. Rodan-Legrain, J.M. Park, N.F. Yuan, K. Watanabe, T. Taniguchi, R.M. Fernandes, L. Fu, P. Jarillo-Herrero, Nematicity and competing orders in superconducting magic-angle graphene, Science 372 (2021) 264-271, https://doi.org/10.1126/science.abc283.

.

[24] H.C. Po, L. Zou, A. Vishwanath, T. Senthil, Origin of Mott insulating behavior and superconductivity in twisted bilayer graphene, Physical Review X 8(3) (2018) 031089, https://doi.org/10.1103/PhysRevX.8.031089.

[25] M. Yankowitz, S. Chen, H. Polshyn, Y. Zhang, K. Watanabe, T. Taniguchi, D. Graf, A.F. Young, C.R. Dean, Tuning superconductivity in twisted bilayer graphene, Science 363 (2019) 1059-1064, https://doi.org/10.1126/science.aav1910.

[26] G. Tarnopolsky, A.J. Kruchkov, A. Vishwanath, Origin of magic angles in twisted bilayer graphene, Phy. Rev. Let. 122(10) (2019)106405, https://doi.org/10.1103/PhysRevLett.122.106405.

[27] M. Koshino, N.N. Nam, Effective continuum model for relaxed twisted bilayer graphene and moiré electron-phonon interaction, Phy. Rev. B 101(19) (2020) 195425, https://doi.org/10.1103/PhysRevB.101.195425.

[28] Y.W. Choi, H.J. Choi, Dichotomy of electron-phonon coupling in graphene moire flat bands, Physical Review Letters 127(16) (2021) 167001, https://doi.org/10.1103/PhysRevLett.127.167001.

[29] Y.W. Choi, H.J. Choi, Strong electron-phonon coupling, electron-hole asymmetry, and nonadiabaticity in magic-angle twisted bilayer graphene, Phy. Rev. B 98(24) (2018) 241412, https://doi.org/10.1103/PhysRevB.98.241412.

[30] F. Wu, A.H. MacDonald, I. Martin, Theory of phonon-mediated superconductivity in twisted bilayer graphene, Physical Review Letters 121(25) (2018), 257001, https://doi.org/10.1103/PhysRevLett.121.257001.

[31] B. Lian, Z. Wang, B.A. Bernevig, Twisted bilayer graphene: a phonon-driven superconductor, Physical Review Letters 122(25) (2019) 257002, https://doi.org/10.1103/PhysRevLett.122.257002.

[32] X. Liu, R. Peng, Z. Sun, J. Liu, Moiré phonons in magic-angle twisted bilayer graphene, Nano letters 22(19) (2022) 7791-7797, https://doi.org/10.1021/acs.nanolett.2c02010.

[33] M. Koshino, Y.-W. Son, Moiré phonons in twisted bilayer graphene, Physical Review B 100(7) (2019) 075416, https://doi.org/10.1103/PhysRevB.100.075416.

[34] A.I. Cocemasov, D.L. Nika, A.A. Balandin, Phonons in twisted bilayer graphene, Physical Review B 88(3) (2013) 035428, https://doi.org/10.1103/PhysRevB.88.035428.

[35] Z. Li, J.-M. Lai, J. Zhang, Review of phonons in moiré superlattices, Journal of Semiconductors 44(1) (2023) 011902, https://doi.org/10.1088/1674-4926/44/1/011902.

[36] D.L. Nika, A.A. Balandin, Phonons and thermal transport in graphene and graphene-based materials, Reports on Progress in Physics 80(3) (2017) 036502, https://doi.org/10.1088/1361-6633/80/3/036502.

[37] M. Lamparski, B. Van Troeye, V. Meunier, Soliton signature in the phonon spectrum of twisted bilayer graphene, 2D Materials 7(2) (2020) 025050, https://doi.org/10.1088/2053-1583/ab7874.

[38] H. Hao, M.-L. Lin, B. Xu, H. Wu, Y. Wang, H. Peng, P.-H. Tan, L. Tong, J. Zhang, Enhanced Layer-Breathing Modes in van der Waals Heterostructures Based on Twisted Bilayer Graphene, ACS nano (2023), https://doi.org/10.1021/acsnano.3c00022.

[39] A. Schäpers, J. Sonntag, L. Valerius, B. Pestka, J. Strasdas, K. Watanabe, T. Taniguchi, L. Wirtz, M. Morgenstern, B. Beschoten, Raman imaging of twist angle variations in twisted bilayer graphene at intermediate angles, 2D Materials 9(4) (2022), 045009, https://doi.org/10.1088/2053-1583/ac7e59.

[40] T.C. Barbosa, A.C. Gadelha, D.A. Ohlberg, K. Watanabe, T. Taniguchi, G. Medeiros-Ribeiro, A. Jorio, L.C. Campos, Raman spectra of twisted bilayer graphene close to the magic angle, 2D Materials 9(2), (2022) 025007, https://doi.org/10.1088/2053-1583/ac4af9.

.

[41] A. Righi, P. Venezuela, H. Chacham, S. Costa, C. Fantini, R. Ruoff, L. Colombo, W. Bacsa, M. Pimenta, Resonance Raman spectroscopy in twisted bilayer graphene, Solid state communications 175 (2013) 13-17, https://doi.org/10.1016/j.ssc.2013.05.015.

[42] J. Campos-Delgado, L.G. Cançado, C.A. Achete, A. Jorio, J.-P. Raskin, Raman scattering study of the phonon dispersion in twisted bilayer graphene, Nano Research 6 (2013), 269-274, https://doi.org/10.1007/s12274-013-0304-z.

[43] V. Carozo, C.M. Almeida, E.H. Ferreira, L.G. Cancado, C.A. Achete, A. Jorio, Raman signature of graphene superlattices, Nano Letters 11(11) (2011), 4527-4534, https://doi.org/ 10.1021/nl201370m.

[44] Y. Wang, Z. Su, W. Wu, S. Nie, N. Xie, H. Gong, Y. Guo, J. Hwan Lee, S. Xing, X. Lu, Resonance Raman spectroscopy of G-line and folded phonons in twisted bilayer graphene with large rotation angles, Applied Physics Letters 103(12) 2013, http://dx.doi.org/10.1063/1.4821434.

[45] L.T. Kong, Phonon dispersion measured directly from molecular dynamics simulations, Computer Physics Communications 182(10) (2011) 2201-2207, https://doi.org/10.1016/j.cpc.2011.04.019.

[46] A.P. Thompson, H.M. Aktulga, R. Berger, D.S. Bolintineanu, W.M. Brown, P.S. Crozier, P.J. in't Veld, A. Kohlmeyer, S.G. Moore, T.D. Nguyen, LAMMPS-a flexible simulation tool for particle-based materials modeling at the atomic, meso, and continuum scales, Computer Physics Communications 271 (2022) 108171, https://doi.org/10.1016/j.cpc.2021.108171.

[47] C. Hudon, R. Meyer, L.J. Lewis, Low-frequency vibrational properties of nanocrystalline materials: Molecular dynamics simulations of two-dimensional systems, Phy. Rev. B 76(4) (2007) 045409, https://doi.org/10.1103/PhysRevB.76.045409.

[48] Z. Tang, N. Aluru, Calculation of thermodynamic and mechanical properties of silicon nanostructures using the local phonon density of states, Phy. Rev. B 74(23) (2006) 235441, https://doi.org/10.1103/PhysRevB.74.235441.

[49] J.W. Jiang, B.S. Wang, T. Rabczuk, Acoustic and breathing phonon modes in bilayer graphene with Moiré patterns, Applied Physics Letters, 101 (2012) 023113, https://doi.org/10.1063/1.4735246

[50] M. Wen, E.B. Tadmor, Hybrid neural network potential for multilayer graphene, Physical Review B 100(19) (2019) 195419, https://doi.org/10.1103/PhysRevB.100.195419.

[51] S. Ahmed, S. Alam, A. Jain, Understanding phonon thermal transport in twisted bilayer graphene, Physical Review B 108(23) (2023) 235202, https://doi.org/10.1103/PhysRevB.108.235202.

[52] T. Tanimoto, International Business Machines Corporation, New York, (1958).

[53] K.P. Murphy, Machine learning: a probabilistic perspective, MIT press(2012).

[54] P. Milton, H. Coupland, E. Giorgi, S. Bhatt, Spatial analysis made easy with linear regression and kernels, Epidemics, 29 (2019) 100362, https://doi.org/10.1016/j.epidem.2019.100362.

[55] F. Pedregosa, G. Varoquaux, A. Gramfort, V. Michel, B. Thirion, O. Grisel, M. Blondel, P. Prettenhofer, R. Weiss, V. Dubourg, Scikit-learn: Machine learning in Python, the Journal of machine Learning research, 12 (2011) 2825-2830.

.

# Supplemental Information for "Emergent Atomic Environments in Twisted Bilayer Graphene and Their Use in the Prediction of the Vibrational Properties"

Dilara Ickecan[1,2‡], Yunus Emre Okyayli[3], Erdi Ata Bleda[1], Dogan Erbahar[4]

[1]*Marmara University, Dept. of Physics, 34722, Ziverbey, Istanbul, Türkiye*

[2]*Istanbul Ticaret University, Dept. of Computer Programming, 34840, Istanbul, Türkiye*

[3]*Dogus University, Dept. of Mechanical Engineering, Erbahar Researcher Lab, Umraniye, 34775, Istanbul, Türkiye*

[4]*Dogus University, Faculty of Engineering, Dept. of Mechanical Engineering, Umraniye, 34775, Istanbul, Türkiye*

## MACHINE LEARNING MODEL DETAILS

Kernel Ridge Regression (KRR) [1] which was used in this study is a sophisticated machine learning method that combines ridge regression with kernel trick to enhance prediction accuracy in complex datasets. The KRR method is particularly advantageous in scenarios where the relationship between the variables is nonlinear. We opt for the Laplacian kernel, renowned for its efficacy in capturing locality and intricate structures in data. The Laplacian kernel, characterized by its exponential decay, and is particularly suitable for datasets where identifying the local structure is crucial. The alpha parameter in the KRR was set to 0.01, a tuned choice aimed at achieving an optimal balance between model complexity and training data fit. This low alpha value suggests a strong regularization effect, mitigating the risk of overfitting while ensuring sufficient model flexibility to capture the underlying data patterns. The combination of the Laplacian kernel with this specific alpha setting in KRR underpins the robustness and accuracy of our predictive model, particularly in handling high-dimensional and complex data structures.

The use of mean squared error (MSE) in KRR is essential for evaluating model performance, particularly in contexts where multicollinearity and overfitting are concerns. MSE is defined as the square of the average squared difference between the observed and predicted values, mathematically expressed as:

$$MSE = \frac{1}{\mathrm{n}} \sum (\mathrm{y_i} - \mathrm{y'_i})^2$$

where n is the number of observations, $\mathrm{y_i}$ represents observed values, $\mathrm{y'_i}$ denotes predicted values. This statistic is vital for evaluating the predictive strength of our model, indicating how well future data points are likely to be predicted. Concurrently, the mean absolute error (MAE) is utilized to measure the accuracy of predictions. MAE is calculated as the average of the absolute differences between the predicted and observed values, defined as:

$$\mathrm{MAE} = \frac{1}{\mathrm{n}} \sum (|\mathrm{y_i} - \mathrm{y'_i}|)$$

where n is the number of observations. MAE provides a straightforward metric to understand the average error magnitude in predictions, reflecting the typical deviation of predicted values from actual observations. Together, MSE and MAE are instrumental in illustrating the efficiency and accuracy of our model in representing the physical phenomena under investigation.

---

‡ Corresponding Author.

E-mail address: dickecan@ticaret.edu.tr (Dilara Ickecan)

.

# THE TBLG STRUCTURES USED IN THE STUDY

Table S1. The list of structures included in the study with their nomenclature indices, lattice parameters, twisting angles, number of atoms, number of different atom types in the unit cell, their respective ratios and wallpaper groups.

| Moiré Lattice Indices§ | Lattice Parameter | Twisting Angle | Number of Atoms in Unit Cell (N) | Number of Different Atom Types (T) | N/T ratio | Wallpaper Group |
|---|---|---|---|---|---|---|
| [1, 1], [1, 0] | 2.48 | 60 | 4 | 2 | 2,00 | p31m |
| [3, 2], [3, 1] | 6.56 | 21.78 | 28 | 6 | 4,67 | p6m |
| [4, 3], [4, 1] | 8.94 | 32.20 | 52 | 10 | 5,20 | p6m |
| [5, 3], [5, 2] | 10.81 | 13.17 | 76 | 14 | 5,43 | p6m |
| [5, 4], [5, 1] | 11.36 | 38.21 | 84 | 3 | 28,00 | p31m |
| [6, 5], [6, 1] | 13.81 | 42.10 | 124 | 22 | 5,64 | p6m |
| [7, 4], [7, 3] | 15.09 | 9.43 | 148 | 27 | 5,48 | p6m |
| [7, 5], [7, 2] | 15.49 | 27.79 | 156 | 5 | 31,20 | p31m |
| [7, 6], [7, 1] | 16.26 | 44.82 | 172 | 31 | 5,55 | p6m |
| [8, 5], [8, 3] | 17.36 | 16.42 | 196 | 34 | 5,76 | p6m |
| [8, 7], [8, 1] | 18.72 | 46.83 | 228 | 7 | 32,57 | p31m |
| [9, 5], [9, 4] | 19.37 | 7.34 | 244 | 42 | 5,81 | p6m |
| [9, 7], [9, 2] | 20.30 | 35.57 | 268 | 47 | 5,70 | p6m |
| [9, 8], [9, 1] | 21.19 | 48.36 | 292 | 50 | 5,84 | p6m |
| [10, 7], [10,3] | 22.04 | 26.01 | 316 | 54 | 5,85 | p6m |
| [11, 6], [11,5] | 23.66 | 6.01 | 364 | 62 | 5,87 | p6m |
| [10, 9], [10, 1] | 23.66 | 49.58 | 364 | 62 | 5,87 | p6m |
| [11, 7], [11, 4] | 23.92 | 17.90 | 372 | 11 | 33,82 | p31m |
| [11, 8], [11, 3] | 24.43 | 29.41 | 388 | 66 | 5,88 | p6m |
| [11, 9], [11, 2] | 25.17 | 40.35 | 412 | 70 | 5,89 | p6m |
| [12, 7], [12, 5] | 25.89 | 10.99 | 436 | 75 | 5,81 | p6m |
| [11, 10],[11, 1] | 26.13 | 50.57 | 444 | 13 | 34,15 | p31m |
| [13, 7], [13, 6] | 27.95 | 5.09 | 508 | 86 | 5,91 | p6m |
| [13, 8], [13, 5] | 28.17 | 15.18 | 516 | 15 | 34,40 | p31m |
| [13, 9], [13, 4] | 28.60 | 25.04 | 532 | 90 | 5,91 | p6m |
| [12, 11], [12,1] | 28.60 | 51.39 | 532 | 94 | 5,66 | p6m |
| [13, 10], [13,3] | 29.24 | 34.54 | 556 | 94 | 5,91 | p6m |
| [13, 11], [13,2] | 30.07 | 43.57 | 588 | 17 | 34,59 | p31m |
| [14, 9], [14,5] | 30.47 | 18.73 | 604 | 103 | 5,86 | p6m |
| [13, 12], [13,1] | 31.07 | 52.07 | 628 | 106 | 5,92 | p6m |
| [14, 11], [14,3] | 31.66 | 36.52 | 652 | 110 | 5,93 | p6m |
| [15, 8], [15,7] | 32.24 | 4.41 | 676 | 114 | 5,93 | p6m |
| [15, 11], [15,4] | 33.36 | 30.16 | 724 | 123 | 5,89 | p6m |
| [14, 13], [14, 1] | 33.55 | 52.66 | 732 | 21 | 34,86 | p31m |
| [16, 9], [16, 7] | 34.45 | 8.26 | 772 | 134 | 5,76 | p6m |
| [15, 13], [15, 2] | 34.98 | 45.89 | 796 | 136 | 5,85 | p6m |
| [16, 11], [16, 5] | 35.16 | 24.43 | 804 | 24 | 33,50 | p31m |
| [15, 14], [15, 1] | 36.024 | 53.16 | 844 | 144 | 5,86 | p6m |
| [17, 9], [17, 8] | 36.533 | 3.89 | 868 | 148 | 5,86 | p6m |
| [16, 13], [16, 3] | 36.533 | 39.68 | 868 | 146 | 5,95 | p6m |
| [17, 10], [17, 7] | 36.701 | 11.63 | 876 | 25 | 35,04 | p31m |
| [17, 11], [17, 6] | 37.034 | 19.27 | 892 | 150 | 5,95 | p6m |
| [17, 12], [17, 5] | 37.529 | 26.74 | 916 | 154 | 5,95 | p6m |
| [17, 13], [17, 4] | 38.179 | 33.99 | 948 | 27 | 35,11 | p31m |
| [16, 15], [16, 1] | 38.500 | 53.60 | 964 | 166 | 5,81 | p6m |
| [18, 11], [18, 7] | 38.976 | 14.62 | 988 | 169 | 5,85 | p6m |
| [17, 14], [17, 3] | 38.976 | 40.97 | 988 | 166 | 5,95 | p6m |
| [18, 13], [18, 5] | 39.912 | 28.78 | 1036 | 175 | 5,92 | p6m |
| [17, 15], [17, 2] | 39.912 | 47.64 | 1036 | 176 | 5,89 | p6m |
| [19, 10], [19, 9] | 40.826 | 3.48 | 1084 | 182 | 5,96 | p6m |
| [19, 11], [19, 8] | 40.976 | 10.42 | 1092 | 31 | 35,23 | p31m |

§ Defined by Ref. [1] A. Artaud, L. Magaud, T. Le Quang, V. Guisset, P. David, C. Chapelier, J. Coraux, Universal classification of twisted, strained and sheared graphene moiré superlattices, Scientific reports, 6 (2016) 25670.

.

| | | | | | | |
|---|---|---|---|---|---|---|
| [17, 16], [17, 1] | 40.976 | 53.99 | 1092 | 31 | 35,23 | p31m |
| [19, 12], [19, 7] | 41.275 | 17.28 | 1108 | 187 | 5,93 | p6m |
| [19, 13], [19, 6] | 41.720 | 24.02 | 1132 | 190 | 5,96 | p6m |
| [19, 14], [19, 5] | 42.306 | 30.59 | 1164 | 33 | 35,27 | p31m |
| [20, 11], [20, 9] | 43.026 | 6.61 | 1204 | 202 | 5,96 | p6m |
| [19, 15], [19, 4] | 43.026 | 36.96 | 1204 | 202 | 5,96 | p6m |
| [18, 17], [18, 1] | 43.453 | 54.33 | 1228 | 206 | 5,96 | p6m |
| [20, 13], [20, 7] | 43.594 | 19.65 | 1236 | 35 | 35,31 | p31m |
| [19, 16], [19, 3] | 43.876 | 43.11 | 1252 | 211 | 5,93 | p6m |
| [19, 17], [19, 2] | 44.846 | 49.00 | 1308 | 38 | 34,42 | p31m |
| [21, 11], [21,10] | 45.120 | 3.15 | 1324 | 224 | 5,91 | p6m |
| [21, 13], [21, 8] | 45.527 | 15.65 | 1348 | 227 | 5,94 | p6m |
| [19, 18], [19, 1] | 45.930 | 54.64 | 1372 | 232 | 5,91 | p6m |
| [20, 17], [20, 3] | 46.330 | 44.01 | 1396 | 236 | 5,92 | p6m |
| [21, 16], [21, 5] | 47.120 | 33.65 | 1444 | 243 | 5,94 | p6m |
| [22, 13], [22, 9] | 47.510 | 11.98 | 1468 | 246 | 5,97 | p6m |
| [21, 17], [21, 4] | 47.897 | 39.33 | 1492 | 252 | 5,92 | p6m |
| [22, 15], [22, 7] | 48.280 | 23.71 | 1516 | 254 | 5,97 | p6m |
| [20, 19], [20, 1] | 48.408 | 54.91 | 1524 | 43 | 35,44 | p31m |
| [23, 12], [23,11] | 49.414 | 2.89 | 1588 | 268 | 5,93 | p6m |
| [23, 13], [23,10] | 49.538 | 8.61 | 1596 | 47 | 33,96 | p31m |
| [22, 17], [22, 5] | 49,538 | 34.96 | 1596 | 45 | 35,47 | p31m |
| [23, 14], [23, 9] | 49,786 | 14.31 | 1612 | 273 | 5,90 | p6m |
| [21, 19], [21, 2] | 49,786 | 50.10 | 1612 | 277 | 5,82 | p6m |
| [23, 15], [23, 8] | 50,155 | 19.93 | 1636 | 276 | 5,93 | p6m |
| [23, 16], [23, 7] | 50.643 | 25.46 | 1668 | 47 | 35,49 | p31m |
| [21, 20], [21, 1] | 50.885 | 55.16 | 1684 | 286 | 5,89 | p6m |
| [23, 17], [23, 6] | 51.247 | 30.87 | 1708 | 286 | 5,97 | p6m |
| [22, 19], [22, 3] | 51.247 | 45.55 | 1708 | 286 | 5,97 | p6m |
| [24, 13], [24,11] | 51.605 | 5.51 | 1732 | 291 | 5,95 | p6m |
| [23, 18], [23, 5] | 51.962 | 36.15 | 1756 | 294 | 5,97 | p6m |
| [23, 19], [23, 4] | 52.784 | 41.27 | 1812 | 52 | 34,85 | p31m |
| [24, 17], [24, 7] | 53.016 | 27.05 | 1828 | 308 | 5,94 | p6m |
| [22, 21], [22, 1] | 53.363 | 55.39 | 1852 | 314 | 5,90 | p6m |
| [25, 13], [25,12] | 53.708 | 2.65 | 1876 | 317 | 5,92 | p6m |
| [23, 20], [23, 3] | 53.708 | 46.22 | 1876 | 320 | 5,86 | p6m |
| [25, 14], [25,11] | 53.822 | 7.93 | 1884 | 53 | 35,55 | p31m |
| [25, 16], [25, 9] | 54.391 | 18.36 | 1924 | 325 | 5,92 | p6m |
| [24, 19], [24, 5] | 54.391 | 37.22 | 1924 | 323 | 5,96 | p6m |
| [23, 21], [23, 2] | 54.729 | 50.99 | 1948 | 327 | 5,96 | p6m |
| [25, 17], [25, 8] | 54.841 | 23.48 | 1956 | 55 | 35,56 | p31m |
| [25, 18], [25, 7] | 55.399 | 28.51 | 1996 | 334 | 5,98 | p6m |
| [23, 22], [23, 1] | 55.841 | 55.59 | 2028 | 57 | 35,58 | p31m |
| [26, 15], [26,11] | 56.061 | 10.15 | 2044 | 343 | 5,96 | p6m |
| [25, 19], [25, 6] | 56.061 | 33.42 | 2044 | 344 | 5,94 | p6m |
| [26, 17], [26, 9] | 56.716 | 20.14 | 2092 | 350 | 5,98 | p6m |
| [25, 21], [25, 4] | 57.683 | 42.86 | 2164 | 364 | 5,95 | p6m |
| [26, 19], [26, 7] | 57.790 | 29.84 | 2172 | 64 | 33,94 | p31m |
| [27, 14], [27,13] | 58.002 | 2.45 | 2188 | 372 | 5,88 | p6m |
| [27, 16], [27,11] | 58.320 | 12.20 | 2212 | 373 | 5,93 | p6m |
| [24, 23], [24, 1] | 58.320 | 55.78 | 2212 | 372 | 5,95 | p6m |
| [27, 17], [27,10] | 58.635 | 17.03 | 2236 | 375 | 5,96 | p6m |
| [25, 22], [25, 3] | 58.635 | 47.38 | 2236 | 376 | 5,95 | p6m |
| [26, 21], [26, 5] | 59.261 | 39.12 | 2284 | 389 | 5,87 | p6m |
| [27, 19], [27, 8] | 59.572 | 26.47 | 2308 | 387 | 5,96 | p6m |
| [25, 23], [25, 2] | 59.675 | 51.74 | 2316 | 67 | 34,57 | p31m |
| [28, 15], [28,13] | 60.188 | 4.72 | 2356 | 395 | 5,96 | p6m |
| [27, 20], [27, 7] | 60.188 | 31.07 | 2356 | 397 | 5,93 | p6m |
| [28, 17], [28,11] | 60.595 | 14.10 | 2388 | 67 | 35,64 | p31m |
| [25, 24], [25, 1] | 60.798 | 55.95 | 2404 | 403 | 5,97 | p6m |
| [26, 23], [26, 3] | 61.101 | 47.89 | 2428 | 408 | 5,95 | p6m |
| [28, 19], [28, 9] | 61.402 | 23.30 | 2452 | 412 | 5,95 | p6m |

.

| | | | | | | |
|---|---|---|---|---|---|---|
| [27, 22], [27, 5] | 61.702 | 39.95 | 2476 | 415 | 5,97 | p6m |
| [29, 15], [29,14] | 62.297 | 2.28 | 2524 | 428 | 5,90 | p6m |
| [29, 16], [29,13] | 62.396 | 6.84 | 2532 | 73 | 34,68 | p31m |
| [27, 23], [27, 4] | 62.592 | 44.22 | 2548 | 426 | 5,98 | p6m |
| [29, 17], [29,12] | 62.592 | 11.37 | 2548 | 429 | 5,94 | p6m |
| [29, 18], [29,11] | 62.886 | 15.87 | 2572 | 430 | 5,98 | p6m |
| [26, 25], [26, 1] | 63.276 | 56.11 | 2604 | 73 | 35,67 | p31m |
| [29, 19], [29,10] | 63.276 | 20.32 | 2604 | 74 | 35,19 | p31m |
| [29, 20], [29, 9] | 63.761 | 24.70 | 2644 | 443 | 5,97 | p6m |
| [28, 23], [28, 5] | 64.145 | 40.72 | 2676 | 75 | 35,68 | p31m |
| [29, 21], [29, 8] | 64.337 | 29.02 | 2692 | 452 | 5,96 | p6m |
| [27, 25], [27, 2] | 64.623 | 52.38 | 2716 | 455 | 5,97 | p6m |
| [30, 17], [30,13] | 64.623 | 8.80 | 2716 | 456 | 5,96 | p6m |
| [29, 22], [29, 7] | 65.002 | 33.25 | 2748 | 79 | 34,78 | p31m |
| [30, 19], [30,11] | 65.191 | 17.50 | 2764 | 464 | 5,96 | p6m |
| [27, 26], [27, 1] | 65.755 | 56.26 | 2812 | 472 | 5,96 | p6m |
| [29, 23], [29, 6] | 65.755 | 37.40 | 2812 | 470 | 5,98 | p6m |
| [28, 25], [28, 3] | 66.035 | 48.80 | 2836 | 475 | 5,97 | p6m |
| [31, 16], [31,15] | 66.592 | 2.13 | 2884 | 486 | 5,93 | p6m |
| [29, 24], [29, 5] | 66.592 | 41.44 | 2884 | 482 | 5,98 | p6m |
| [31, 17], [31,14] | 66.68 | 6.39 | 2892 | 85 | 34,02 | p31m |
| [31, 18], [31,13] | 66.87 | 10.64 | 2908 | 487 | 5,97 | p6m |
| [31, 19], [31,12] | 67.14 | 14.85 | 2932 | 492 | 5,96 | p6m |
| [30, 23], [30, 7] | 67.42 | 34.23 | 2956 | 496 | 5,96 | p6m |
| [29, 25], [29, 4] | 67.51 | 45.38 | 2964 | 83 | 35,71 | p31m |
| [33, 17],[33,16] | 70.87 | 2.00 | 3268 | 547 | 5,97 | p6m |
| [35, 18],[35,17] | 75.18 | 1.89 | 3676 | 625 | 5,88 | p6m |
| [37, 19],[37,18] | 79.48 | 1.79 | 4108 | 692 | 5,94 | p6m |

## SCREE PLOTS

The scree plots of the principal components to explain the variances of the data coming from all structures are represented below.

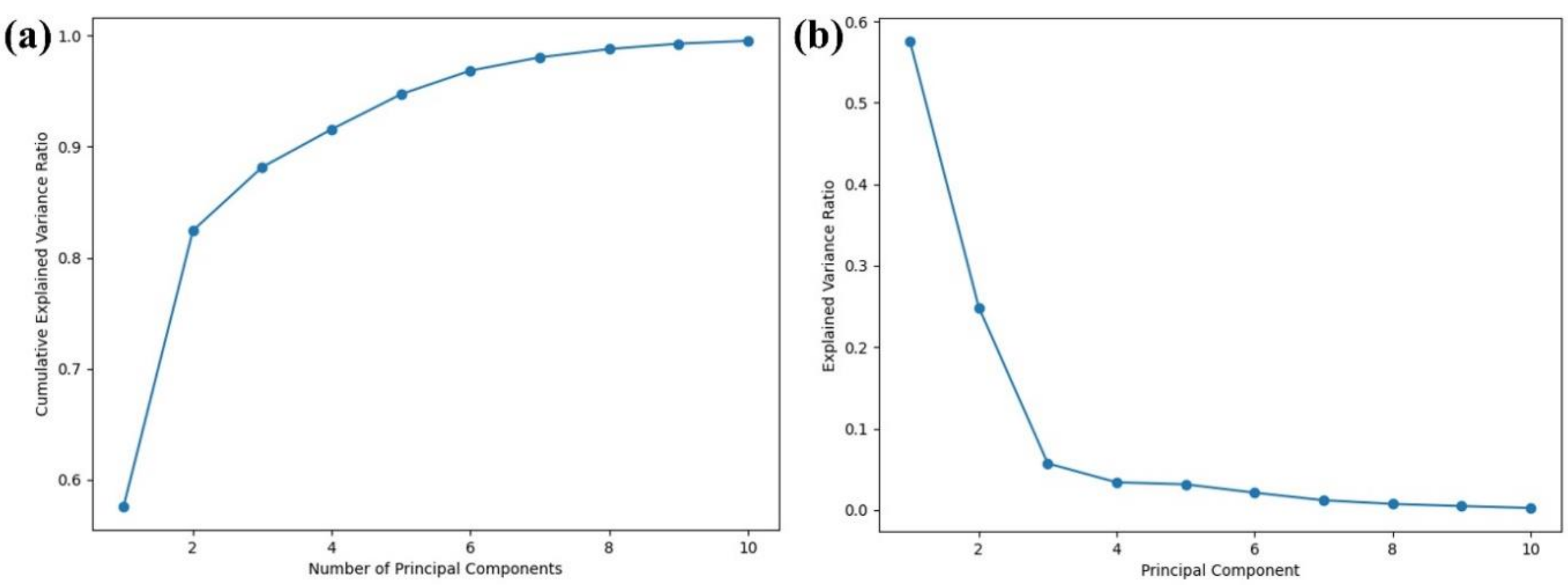

Figure S1. (Color online) (a) A scree plot that shows the % percentage of the cumulative explained variance in the data. The first 6 principal components explain 95% of the total variation. (b) Variances explained by first 10 principal components.

.

## TOTAL PHONON DENSITY OF STATES OF MAGIC-ANGLE TBLG

The magic-angle graphene which has a rotation angle of 1.08° has 11164 atoms in the unit cell. It has 1862 different type of local environments and belongs to the p31m space group. This huge structure doesn't belong to our training and data set however its PTDOS can be predicted from the tuned parameters of our ML model as shown in Figure S2.

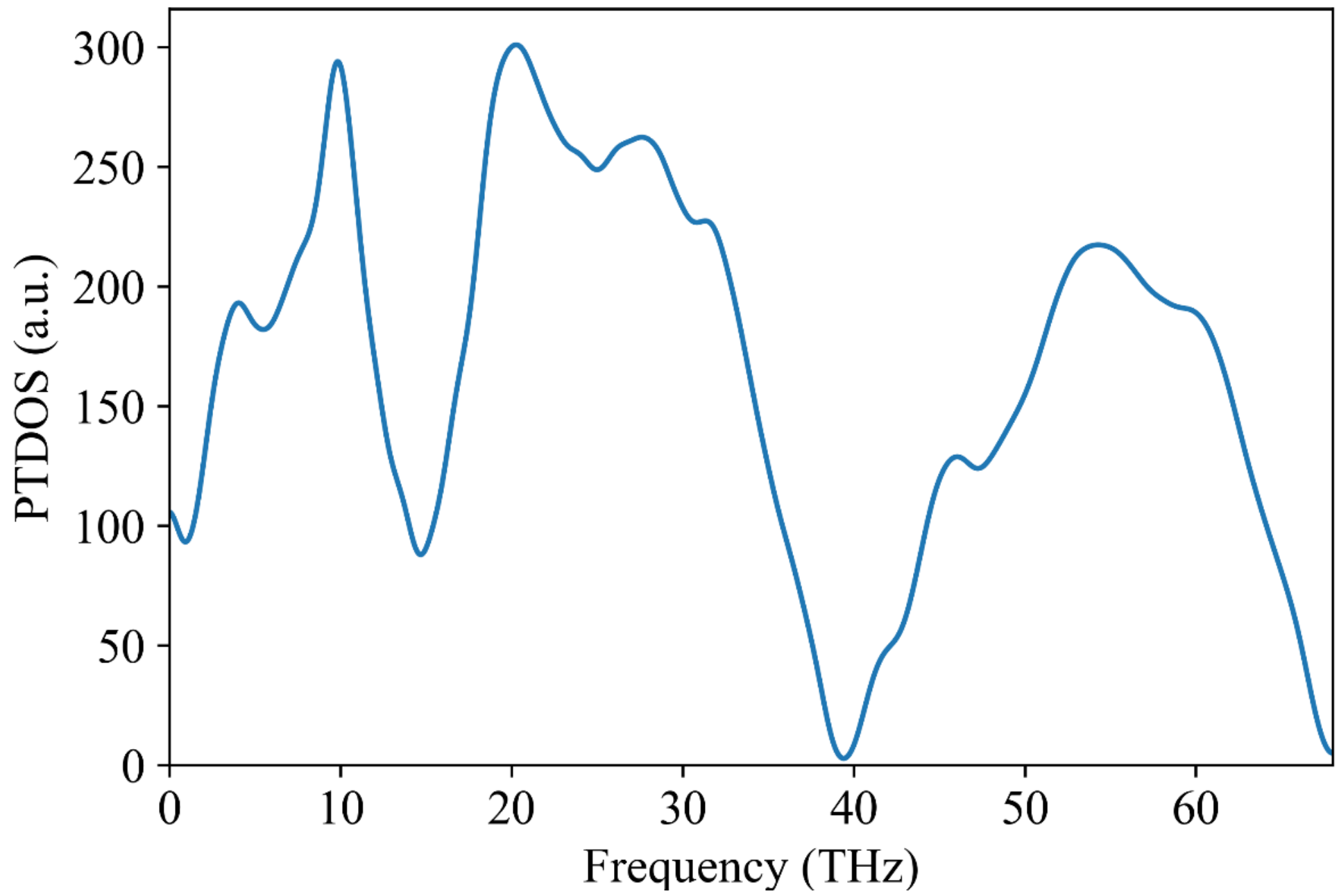

Figure S2. (Color online) (a) PTDOS of 1.08° rotated TBLG predicted from trained model.

## RESULTS OF THE OTHER 4 RANDOMIZATION SEEDS

In the main text Figure 6 shows the results of the best performing seed. Here in Figure S3 we demonstrate the results for the other 4 seeds. A seed initiates the process of randomly choosing 112 structures for training set and leaving the remaining 28 for the test set. After the identification of unique atomic environments in the chosen structures, 50 % of these are selected randomly to train the model by Scikit-learn [3]. The structures in the training and test sets produce different number of atom types for each seed. These numbers are shown in Table S2.

Table S2. Number of identified atomic environments emerging from different randomization seeds. 50 % of the identified environments was chosen randomly to train the model. The number of different environments in the structures that were reserved for test set is shown in the last column.

| | **Identified** | **Training** | **Test** |
|---|---|---|---|
| SEED 1 | 27576 | 13788 | 6818 |
| SEED 2 | 28583 | 14291 | 5811 |
| SEED 3 | 27600 | 13800 | 6794 |
| SEED 4 | 27298 | 13649 | 7096 |

.

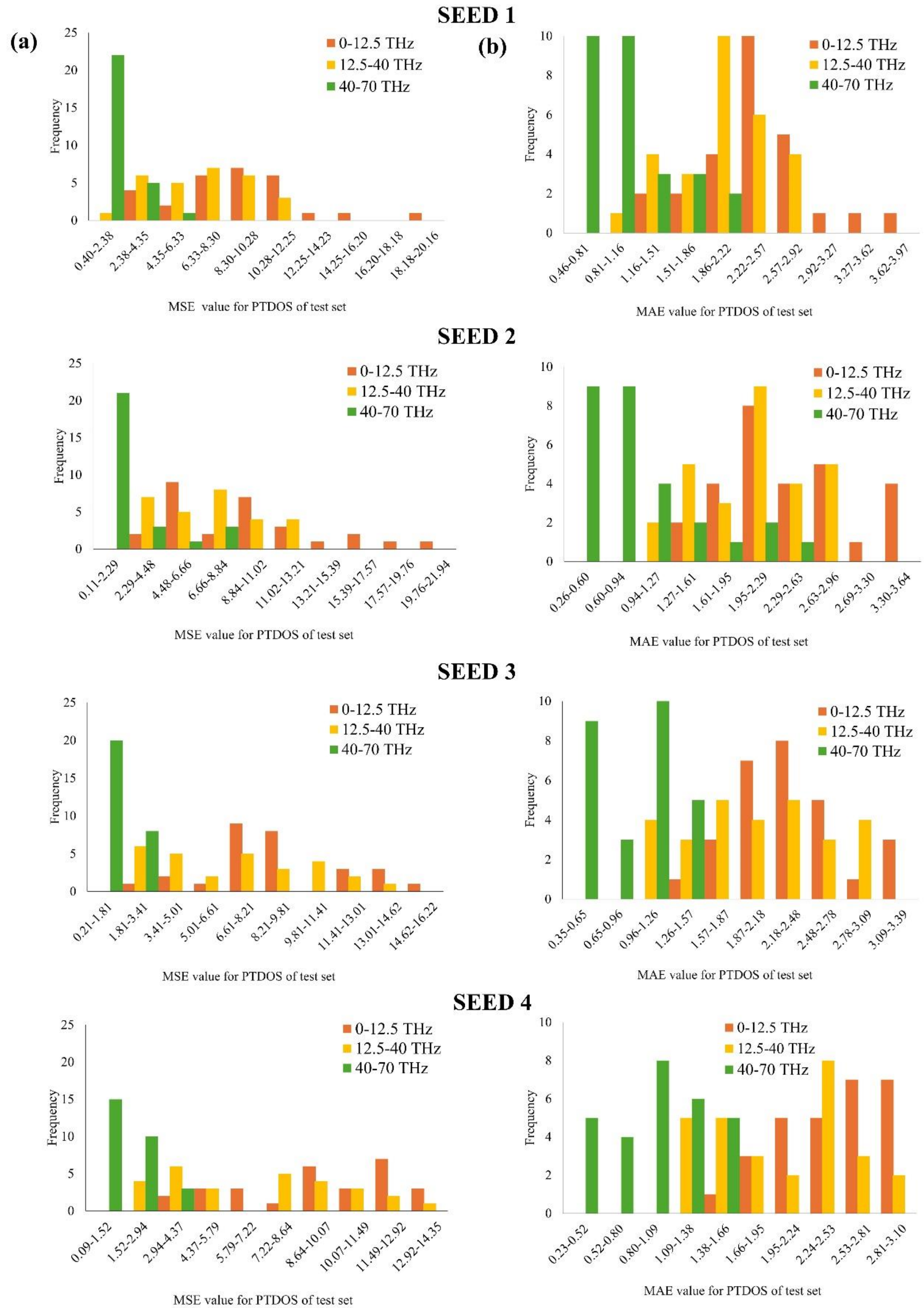

Figure S3. (Color online) Histograms of MSE (a) and MAE (b) values of predicted PTDOS of 28 TBLG structures in the test set for the respective seeds. The histograms are calculated for three regions of the spectra.

.

All MSE and MAE values for the predicted PTDOS of 28 TBLG structures in the test set for other seeds are also shown in Figure S2(a)-(b) as histograms.

The mean values of the MSE and MAE of the 28 structures in the test set for the respective seeds are reported in Table S3.

Table S3. The mean values of MSE and MAE of 28 TBLG structures used in test set for the respective seeds.

| | **SEED 1** | | **SEED 2** | | **SEED 3** | | **SEED4** | |
|---|---|---|---|---|---|---|---|---|
| **Range (THz)** | **MSE** | **MAE** | **MSE** | **MAE** | **MSE** | **MAE** | **MSE** | **MAE** |
| 0 – 12.5 | 9.21 | 2.40 | 9.47 | 2.41 | 8.86 | 2.37 | 9.49 | 2.47 |
| 12.5 – 40 | 6.76 | 2.04 | 7.18 | 2.08 | 6.77 | 2.05 | 6.82 | 2.01 |
| 40 – 70 | 1.79 | 1.03 | 1.85 | 0.96 | 1.41 | 0.91 | 1.60 | 0.98 |

The mean values of the MAE and MSE values for all 5 seeds are given in Table S4.

Table S4. The mean values of MSE and MAE of all 5 seeds.

| **Range (THz)** | **MSE** | **MAE** |
|---|---|---|
| 0 – 12.5 | 9.05 | 2.40 |
| 12.5 – 40 | 6.78 | 2.03 |
| 40 – 70 | 1.65 | 0.97 |

**References**

[1] K.P. Murphy, Machine learning: a probabilistic perspective, MIT press2012.
[2] A. Artaud, L. Magaud, T. Le Quang, V. Guisset, P. David, C. Chapelier, J. Coraux, Universal classification of twisted, strained and sheared graphene moiré superlattices, Scientific reports, 6 (2016) 25670.
[3] F. Pedregosa, G. Varoquaux, A. Gramfort, V. Michel, B. Thirion, O. Grisel, M. Blondel, P. Prettenhofer, R. Weiss, V. Dubourg, Scikit-learn: Machine learning in Python, the Journal of machine Learning research, 12 (2011) 2825-2830.

.